\documentclass[apjl]{emulateapj}

\usepackage{epsfig}

\usepackage{ulem} % DCE
\usepackage{color} % DCE

\newcount\listnorom
\newcommand\listromanDE{\global\advance \listnorom by 1
{\lowercase\expandafter{(\romannumeral\listnorom)}\ }}
\newcommand\newlistroman{\listnorom=0}
\newcommand{\Iratio}{I_\mathrm{MC}/I_\mathrm{SNR}}
\newcommand{\CD}{contact discontinuity}
\newcommand{\fpp}{f_p(p)}
\newcommand{\fep}{f_e(p)}
\newcommand{\DKol}{D_\mathrm{Kol}}

\newcommand{\Egam}{E_\gamma}
\newcommand{\McloudPP}{`molecular cloud'}
\newcommand{\Mcloud}{molecular cloud}
\newcommand{\SC}{self-consistent}
\newcommand{\SCly}{self-consistently}
\newcommand{\RadFS}{R_\mathrm{FS}}
\newcommand{\gamray}{$\gamma$-ray}

\newcommand{\muG}{$\mu$G}
\newcommand{\pcc}{cm$^{-3}$}
\newcommand{\nH}{n_p}
\newcommand{\nMC}{n_\mathrm{MC}}
\newcommand{\rel}{relativistic}

\newcommand{\rg}{r_g}
\newcommand{\rgz}{r_{g0}}
\newcommand{\MS}{M_S}
\newcommand{\MFA}{magnetic field amplification}
\newcommand{\Rtot}{r_\mathrm{tot}}
\newcommand{\Rsub}{r_\mathrm{sub}}
\newcommand{\rTP}{r_\mathrm{TP}}
\newcommand{\fsk}{f_\mathrm{sk}}
\newcommand{\etamfp}{\eta_\mathrm{mfp}}
\newcommand{\EffRel}{\epsilon_\mathrm{acc}}
\newcommand{\EffCRs}{E_\mathrm{CR}/E_\mathrm{SN}}
\newcommand{\Eesc}{\epsilon_\mathrm{esc}}
\newcommand{\tstep}{t_\mathrm{step}}
\newcommand{\coul}{Coulomb}

\newcommand{\syn}{synchrotron}

\newcommand{\brem}{bremsstrahlung}
\newcommand{\brems}{bremsstrahlung}

\newcommand{\pion}{pion-decay}
\newcommand{\Nshell}{N_\mathrm{shell}}

\newcommand{\alfcut}{\alpha_\mathrm{cut}}
\newcommand{\epRatio}{(e/p)_\mathrm{rel}}
\newcommand{\Tep}{(T_e/T_p)}
\newcommand{\Msun}{\mathrm{M}_{\odot}}
\newcommand{\Bcloud}{B_\mathrm{MC}}
\newcommand{\Bism}{B_\mathrm{ISM}}

\newcommand{\Emax}{E_\mathrm{max}}
\newcommand{\pmax}{p_\mathrm{max}}
\newcommand{\tSNR}{t_\mathrm{SNR}}
\newcommand{\dSNR}{D_\mathrm{SNR}}
\newcommand{\EnSN}{E_\mathrm{SN}}
\newcommand{\Mej}{M_\mathrm{ej}}
\newcommand{\SA}{semi-analytic}
\newcommand{\Inj}{\chi_\mathrm{inj}}

\newcommand{\NL}{nonlinear}
\newcommand{\TP}{test-particle}
\newcommand{\Pcr}{P_\mathrm{cr}}

\newcount\Inum
\newcount\IInum
\newcount\IIInum
\newcount\IVnum
\def\I{\global\multiply\IInum by 0 \global\multiply\IIInum by 0
            \global\multiply\IVnum by 0 \global\advance \Inum by 1
            {\the\Inum. }}
\def\II{\global\multiply\IIInum by 0\global\multiply\IVnum by 0
       \global\advance \IInum by 1 {\the\Inum.\the\IInum. }}
\def\III{\global\multiply\IVnum by 0\global\advance \IIInum by 1
            {\the\Inum.\the\IInum.\the\IIInum. }}
\def\IV{\global\advance \IVnum by 1
            {\the\IVnum. }}
\shorttitle{Broadband Spectrum of SNR}
\shortauthors{Lee et al.}

\begin{document}

\submitted{Accepted by ApJ, June 24, 2008}
%top matter
%\usefont{T1}{ptm}{m}{n}
\title{
%A Broadband Emission Spectrum Modeling 
%       of Shell-Type Supernova Remnants on the Non-Linear Diffusive Shock 
%       Acceleration Theory
3-D Model of Broadband Emission from
Supernova Remnants Undergoing Non-linear Diffusive Shock
Acceleration}
\author{Shiu-Hang Lee, Tuneyoshi Kamae}
\affil{Stanford Linear Acceleration Center and Kavli Institute for 
       Particle Astrophysics and Cosmology, Stanford University, 
       Menlo Park, CA 94025}
\email{kamae@slac.stanford.edu,shia520@slac.stanford.edu}
\and
\author{Donald C. Ellison}
\affil{Department of Physics, North Carolina State University, 
Box 8202, Raleigh, NC 27695}
\email{don\_ellison@ncsu.edu}

%%\date{November 7, 2007}

\begin{abstract}
We present a 3-dimensional model of supernova remnants
  (SNRs) where the hydrodynamical evolution of the remnant is modeled
  consistently with nonlinear diffusive shock acceleration
  occuring at the outer blast wave. The model
  includes particle escape and diffusion outside of the forward shock,
  and particle interactions with arbitrary distributions of external
  ambient material, such as molecular clouds. We
  include synchrotron emission and cooling, bremsstrahlung radiation,
  neutral pion production, inverse-Compton (IC), and Coulomb
  energy-loss. Boardband spectra have been calculated for typical
  parameters including dense regions of gas external to a $1000$ year
  old SNR.
In this paper, we describe the details of our model but do not
attempt a detailed fit to any specific remnant. We also do not include
magnetic field amplification (MFA), even though this effect may be
important in some young remnants. In this first presentation of
the model we don't attempt a detailed fit to any specific remnant.
Our aim is to develop a flexible platform, which  can be
generalized to include effects such as MFA, and which can be easily
adapted to various SNR environments, including Type Ia SNRs,
which explode in a constant density medium, and Type II SNRs, which
explode in a pre-supernova wind.
When applied to a specific SNR, our
model will predict cosmic-ray spectra and multi-wavelength morphology
in projected images for instruments with varying spatial and spectral
resolutions.
We show examples of these spectra and images and emphasize 
the importance of measurements in the hard
X-ray, GeV, and TeV gamma-ray bands for investigating key ingredients in
the acceleration mechanism, and for deducing whether or not TeV emission
is produced by IC from electrons or \pion\ from protons.  
\end{abstract}


\maketitle
%end of top matter

\section{Introduction}

Supernovae (SNe) are the only known sources capable of providing the
energy needed to power the bulk of the galactic cosmic rays (CRs) with
energies below the spectral feature called the ``knee'' around $3 \times
10^{15}$\,eV  \citep[e.g.,][]{Drury83}.
If SNe are the main sources of Galactic CRs, the acceleration mechanism
  must be efficient so that $\gtrsim 10$\% of the total SN explosion
  energy in our Galaxy ends up in cosmic rays
  \citep[e.g.,][]{Hillas2005}.
Observational evidence that the outer blast wave shock accelerates
electrons to ultra-relativistic energies
in some young SNRs \citep[e.g.,][]{KoyamaSN1006_95},
and the existence of a well-developed model of particle acceleration at
shocks, i.e., diffusive shock acceleration (DSA)
\cite[e.g.,][]{Drury83,BE87,JE91} support the above contention.

When confronting observations with theoretical models, however, there
remain a number of important ambiguities and uncertainties from both the
observational and theoretical perspectives.  
Resolution of these ambiguities and uncertainties by new telescopes will
be essential to claim evidence for the pion-decay feature in the
GeV-TeV emission from SNRs.  The Gamma-ray Large Area Space
Telescope (GLAST), to be launched in 2008, will probe this crucial
energy range with unprecedented sensitivity and resolution.

Fundamental questions for CR origin also concern the spectral shape and
maximum ion energy a given SNR can produce. Electron energy spectra
inferred from young SNRs vary and can be 
substantially harder than CR electron spectra observed at Earth, even
after correction for propagation in the galaxy \citep[e.g.,][]{BV2006}.
The maximum CR ion energy SNRs actually produce will remain uncertain
until a firm identification of \pion\ emission is obtained and gamma-ray
emission is detected past a few 100~TeV, the maximum possible electron
energy in SNRs.

There remain other basic questions concerning the DSA mechanism. For
instance, is DSA efficient enough for nonlinear 
effects, such as shock smoothing and magnetic field amplification, 
to become important in young SNRs? 
How does particle injection occur and how does injection and
acceleration vary between electrons and protons? While the
galactic CR electron-to-proton ratio, $\epRatio$, of $0.01$--$0.0025$
observed at Earth at \rel\ energies is often used to constrain the
ratio in SNRs, this ratio has not been observed outside of
the heliosphere.\footnote{We note that while energetic electrons and
protons are observed from solar flares and at low Mach number
heliospheric shocks, these observations provide limited help for
understanding the high Mach number shocks expected in young SNRs and
other astrophysical sourses where a large fraction of the shock energy
is put into \rel\ particles.}
The $\epRatio$ ratio is crucial in deciding whether the $\gamma$-ray
emission from different SNRs, or observed in
different parts of an individual SNR, is of hadronic or leptonic origin.

The recent discovery of spatially thin, hard X-ray filaments in some
young SNRs \citep[e.g.,][]{BambaEtal2003,Uchiyama_J1713_2007} supports previous suggestions
\citep[e.g.,][]{Cowsik80,BL2001,RE92} that the particle acceleration
process can amplify the ambient magnetic field by large factors. If
magnetic field amplification in DSA is as large as now appears to be the
case \citep[e.g.,][]{BKV2003}, it will have far-reaching consequences
not only for understanding the origin of Galactic CRs, but for
interpreting \syn\ emission from shocks throughout the universe.  Since
shocks and related superthermal particle populations exist in diverse
environments, the knowledge gained from studying SNRs will have wide
applicability.

The advent of new space- and ground-based telescopes will result in
observations of SNRs at many different wavelengths with greatly improved
sensitivity and resolution. 
It is even conceivable that features in the CR
spectrum observed at Earth might be associated with nearby SNRs with
future observations \citep[e.g.,][]{EW1999,KobayashiEtal2001}.

In order to take full advantage of current and future observations, and
to improve our understanding of the DSA mechanism, the data must be
analyzed with consistent, broadband photon emission models including
nonlinear effects.
This has prompted us to develop a three-dimensional model of young SNRs
where the evolution of the remnant is coupled to nonlinear diffusive
shock acceleration (NL-DSA) \citep[e.g.,][]{EDB2004,EC2005}, in an
environment with an arbitrary mass distribution. We focus on radiation
from CR electrons and protons and leave the modeling of heavier ions for
future work. 
In this preliminary study, we also ignore other possible
acceleration processes, most notably second-order stochastic
acceleration, and do not include magnetic field amplification.

We believe our work is a significant advance over previous work for
several reasons.  Of particular importance is that we include
``escaping'' particles \SCly. In NL-DSA, a sizeable fraction of the SN
explosion energy can be put into very energetic CRs that escape the
forward shock and stream into the surrounding ISM. These particles will
produce detectable radiation if they interact with dense, external
material.  Another advantage is that we have a ``coherent'' model,
easily expandable to include more complex effects, where the various
environmental and theoretical parameters can be straightforwardly varied
and the resulting radiation can be compared directly with observations.
This is important because all SNe and SNRs are different and complex
with many poorly constrained parameters. It is essential that the
underlying theory consistently model broad-band emission from radio
to TeV $\gamma$-rays taking into account individual characteristics of
the remnants and their environments.

In Sections~\ref{section:DSA} and \ref{section:Diff_Interaction}
we give a brief general description of nonlinear diffusive
shock acceleration and describe the environmental and model
parameters required for a hydrodynamical solution.
We place a time-dependent, spherically symmetric, hydrodynamic
calculation of a SNR, including NL-DSA, in a three-dimensional box
consisting of $51 \times 51 \times 51$ cells.\footnote{The resolution of
the 3-D box is, of course, adjustable and limited only by computational
considerations.}
The energetic particles produced by the outer blast
wave shock propagate through the simulation box where
they interact with an arbitrary distribution of matter placed 
external to the outer shock.
The energetic particles in the box, including those within the SNR,
suffer energy losses and produce broad-band continuum emission spectra
by interacting with the magnetic field, photon field, and matter density
of each cell. In Section~\ref{section:results} we show some
examples including line-of-sight projections of the emitted radiation
which are suitable for comparison with observations.

There are a number of young SNRs under active investigation,
including SNR RX~J1713
\citep[e.g.,][]{AharonianJ1713_2006,Uchiyama_J1713_2007}, Vela
Jr. \citep[e.g.,][]{AharonianVela2005}, RCW~86
\citep[][]{Hoppe_etal_RCW86_2007,Ueno_etal_RCW_2007,RhoEtal2002},
IC~443 \citep[][]{Albert_etal_IC443,Humensky_etalIC443} and
W~28 \citep[][]{AharonianW28}.
However, here we concentrate on a general study using various
parameters typical of young, shell Type Ia SNRs and leave detailed
modeling of individual remnants for future work.

\section{Diffusive Shock Acceleration in SNRs}
\label{section:DSA}
\subsection{The Diffusive Shock Acceleration Theory}
In the \TP\ approximation, diffusive shock acceleration
produces superthermal particles with a power law distribution where the
power-law index depends only on the shock compression ratio, i.e., $f(p)
\propto p^{-\sigma}$, where $\sigma= 3\rTP/(\rTP -1)$, $\rTP$ is the
\TP\ shock compression ratio, $p$ is the particle momentum, and $f(p)$
is the phase-space distribution function \citep[see][ and references
therein]{Drury83,BE87}.
This \TP\ result holds as long as the pressure exerted by the
accelerated particles (i.e., cosmic rays), $\Pcr$, is small compared to
the far upstream momentum flux, $\rho_0 u_0^2$ ($\rho_0$ is the
unshocked density and $u_0$ is the unmodified shock speed). There is
considerable observational evidence, however, that DSA is intrinsically
efficient and shocks with high sonic Mach numbers $\MS \gtrsim 10$ are
expected to accelerate particles efficiently enough that $\Pcr \sim
\rho_0 u_0^2$. In this case, the pressure in accelerated particles feeds
back on the shock structure in a strongly nonlinear fashion
\citep[e.g.,][]{JE91}.

\newlistroman
In NL-DSA, the following effects become important:
\listromanDE a precursor is formed upstream of the viscous subshock with
a length scale comparable to the diffusion length
of the highest momentum particles the shock produces. In the shock
reference frame, the incoming plasma is decelerated and heated in the
precursor before it reaches the subshock;
\listromanDE the production of relativistic particles, and the escape of
some fraction of the highest energy particles from the precursor, soften
the equation of state of the plasma, making the plasma
more compressible and allowing the overall shock
compression ratio to increase, i.e., $\Rtot > \rTP$;
\listromanDE the simple power law of the \TP\ approximation is replaced
by a concave spectrum at superthermal energies. The spectrum is softer than
the \TP\ power law for low momentum particles and harder for high
momentum particles; and
\listromanDE the weak subshock has a compression ratio $\Rsub< \rTP$ so
that the shocked plasma has a lower temperature than would be the case
in the \TP\ approximation \citep[see][ and references therein for
detailed discussions of these effects]{BE99}.  

The modification of the equation-of-state by the production of \rel\
particles and the escaping energy flux in NL-DSA, influences the
evolution of the SNR and numerical approaches have been developed to
describe this process \citep*[e.g.,][]{BEK96a,EDB2004}.
Here we generalize the basic NL-SNR model by including CR propagation
within the remnant and, most importantly, in surrounding material using
a three-dimensional simulation. The escaping particle flux is expected
to dominate interactions outside of the SNR blast wave.

\subsection{CR-Hydro Simulation}
We calculate the hydrodynamic evolution of a SNR with a spherically
symmetric model described in detail in \citet{EPSBG2007} and references
therein (see Fig.~\ref{Boxel_fig}).
The model couples efficient DSA to the hydrodynamics using the \SA\ model
of \citet*{BGV2005} \citep[see also][]{AB2005,AB2006}.
Given an injection parameter, $\Inj$ \citep[this is $\xi$ in
equation~(25) in][]{BGV2005}, the \SA\ model 
  calculates the full proton distribution function $\fpp$ at each
  time-step of the hydro simulation, along with the overall shock
  compression ratio, $\Rtot$, and the subshock compression ratio,
  $\Rsub$. The hydro provides the required input for the \SA\
  calculation, i.e., the shock speed, shock radius, ambient density and
  temperature, and the ambient magnetic field, and $\fpp$ reflects the
  \NL\ effects from efficient acceleration.
The coupling between the hydro and NL-DSA is accomplished by using $\fpp$,
  and the escaping particle flux, to calculate an effective ratio of
  specific heats which is then used in the hydrodynamic equations. The
  electron spectrum, $\fep$, is determined from $\fpp$ with two
  additional parameters, the electron-to-proton ratio at \rel\ energies,
  $\epRatio$, and the temperature ratio immediately behind the shock, $\Tep$
  \citep[see][ for a full discussion]{EDB2004}.

In this paper, we only consider Type Ia supernovae with no pre-SN wind.
We also ignore any CR production that might occur at the reverse
shock. Both of these restrictions are for clarity and our model can
be applied to Type II SNe with winds and can calculate particle heating
and acceleration at reverse shocks.
The parameters controlling our results fall into two catagories:
Environmental parameters and model parameters. These are listed in the
following sections with either default values or the
range of values used for our examples.

\newlistroman
\subsubsection{Environment Parameters}
\label{section:param_envir}
The environmental parameters include:
\listromanDE
the SN explosion energy, $\EnSN = 10^{51}$\,erg,
\listromanDE
the ejecta mass, $\Mej = 1.4\,\Msun$,
\listromanDE
the distance to the SNR, $\dSNR = 1$\,kpc,
\listromanDE
the age of the SNR, $\tSNR = 1000$\,yr,
\listromanDE the ISM proton number density,
$n_p= 0.1, 1,$ or $10$~cm$^{-3}$,
\listromanDE the proton number density in the \Mcloud\ if present
$\nMC=10^3$\,\pcc,
\listromanDE
the ambient, i.e., unshocked, magnetic field, $\Bism = 3$\,\muG, and
\listromanDE
the ambient proton temperature, $T_p = 10^4$~K.
The quantities $n_p$, $\Bism$, and $T_p$ are assumed to be constant in the
region outside of the forward shock.

\newlistroman
\subsubsection{Model Parameters}
\label{section:param_model}
The model parameters used in this simulation are:
\listromanDE
an exponential ejecta density profile applicable to Type Ia SNe,
\listromanDE the acceleration efficiency for DSA, $\EffRel=\EffCRs$,
where we consider two possibilities: the test-particle case where 1\% of
the total SN explosion energy is put into CR energy,
$E_\mathrm{CR}$, during the 1000\,yr evolution of the SNR,
and \NL\ DSA, where 75\% of the SN explosion energy is put into CRs
during 1000\,yr,\footnote{These percentages include CRs that escape
upstream from the forward shock during the SNR evolution.}
\listromanDE
the electron to proton ratio at relativistic energies, 
      $\epRatio = 0.01$,
\listromanDE the electron to proton temperature ratio immediately behind
the forward shock, $\Tep = 1$,
\listromanDE
the cutoff index for the shape of particle spectra near $\Emax$, 
      $\alfcut = 1$,
\listromanDE
the number of gyroradii in a mean free path, $\etamfp=1$,\footnote{This
  parameter is discussed more fully in Section~\ref{section:diff} below.}
\listromanDE
the fraction of the forward shock radius, $\fsk=0.05$, used to truncate
DSA,\footnote{The maximum proton energy produced by the shock,
$\Emax$, is determined by either the finite shock age, $\tSNR$, or
the finite size of the shock, whichever occurs first. Our choice of
$\fsk=0.05$ is arbitrary but is consistent with previous work
\citep[e.g.,][]{EC2005}. For this particular $\fsk$, $\Emax$ is
determined by the finite shock size in all of our
examples.\label{FN:maxsize}}
\listromanDE the number of shells between the forward shock
and the contact discontinuity at the end of the simulation,
$\Nshell = 20$, and
\listromanDE
the diffusive time step interval, $\tstep = 10$\,yr.
All of these parameters, except $\nMC$ and $\tstep$, are described in
detail in \citet{EC2005} and \citet{EPSBG2007}.

The geometry of the magnetic field that is used as input to the DSA
calculation and to calculate the \syn\ emission is not described
explicitly in the CR-hydro simulation. Instead, it is assumed that the
field immediately upstream from the FS, $B_0=\Bism$, is turbulent and,
as in \citet{VBKR2002}, we set the immediate downstream compressed field
to $B_2 = B_0\sqrt{1/3 + 2\Rtot^2/3}$, where $\Rtot$ is the overall
shock compression ratio.
The magnitude of the shocked field evolves as the density of the plasma
changes, as described in \citet{EC2005}, and the magnetic pressure is
included in the hydrodynamics, although it is insignificant for the
results we show here.
An important limitation of our current model is that we do not include
self-generated magnetic turbulence or \MFA. Magnetic field amplification
is only now being studied in nonlinear calculations
\citep[e.g.,][]{AB2006,VEB2006} and we leave implementation of this
important aspect of DSA for future work.  We also neglect other
wave-particle effects, such as wave-damping \citep[e.g.,][]{Pohl2005},
and simply assume that the shocked field is turbulent enough for Bohm
diffusion to occur with a background field that is compressed at the
shock and evolves adiabatically behind the shock.

\subsection{Model Geometry and Simulation Method}
\begin{figure}        % Fig 1
\epsscale{1.0} \plotone{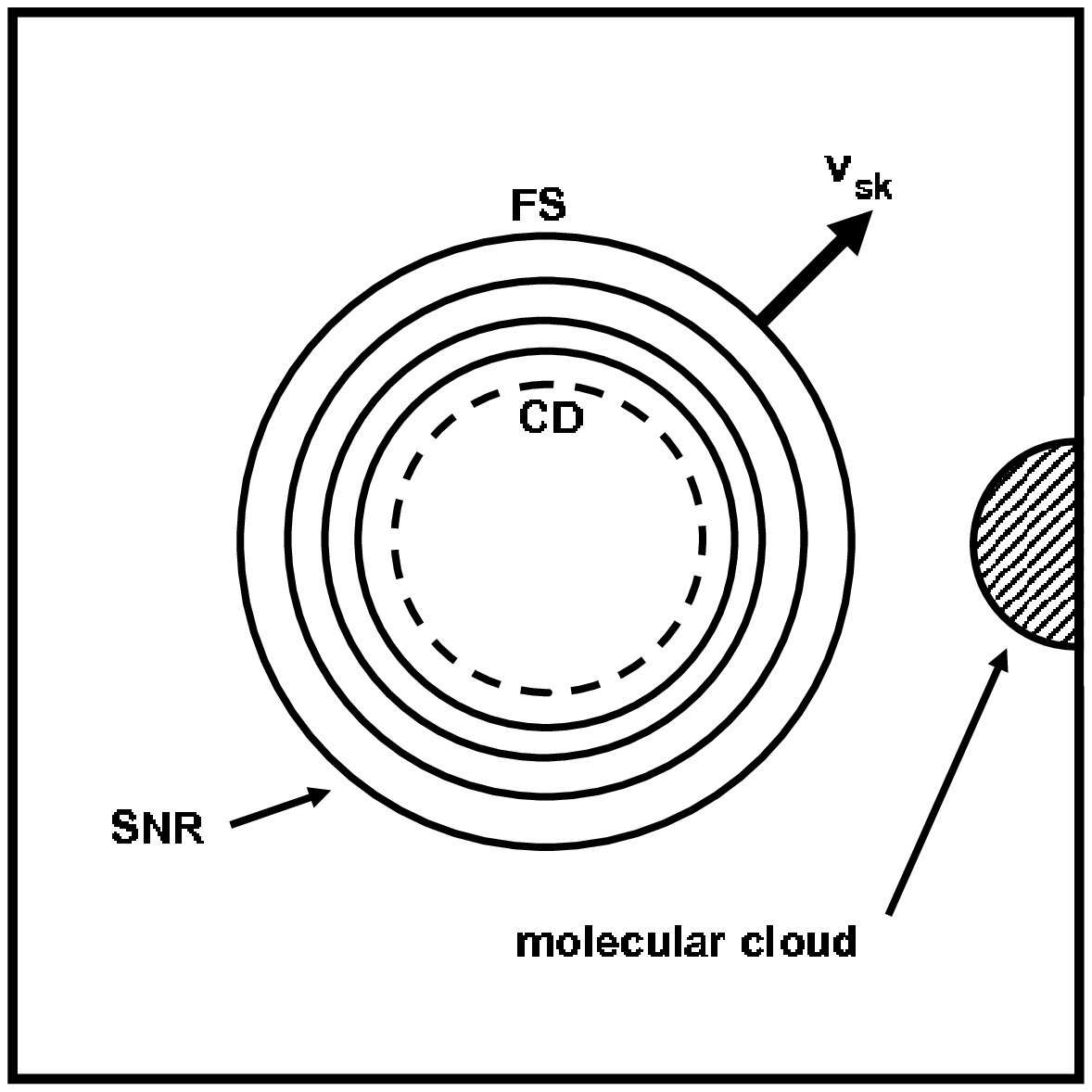}
\caption{Cross-section of the 3-D boxel simulation (not to scale). For
  the results shown in this paper, the box is divided into $51 \times 51
  \times 51$ cells. Only a few representative SNR shells are shown in
  this sketch. The actual number of spherically symmetric shells between
  the forward shock (FS) and \CD (CD) increases with time
  and equals 20 at the end of the simulation. We show the `molecular
  cloud' discussed in Section~\ref{section:profile}. The `molecular
  cloud' shell discussed in Section~\ref{section:MC_shell} is not shown
  for clarity.
\label{Boxel_fig}}
\end{figure}

We treat the SNR hydrodynamics in 1-D by assuming a spherically symmetric
structure for the region of the remnant between the forward shock (FS)
and the contact discontinuity (CD).
The main generalization we have made to the CR-hydro model of
\citet{EPSBG2007} is to imbed the SNR in a fully 3-D astrophysical
environment where CRs accelerated by the remnant propagate and interact
with ambient material. A cross-section of the 3-D simulation box is
shown in Fig.~\ref{Boxel_fig}.
Spatially dependent environmental aspects, like matter density in a
molecular cloud, magnetic field strengths, and the
magnetic turbulence spectrum are all defined and stored in 3-D
simulation cells.

The various interactions and photon emission processes
are computed throughout the simulation box so that mult-wavelength
spectra and projected morphology are obtained.

The temporal sequence of the evolving SNR is as follows:
For a SNR of a given age, $\tSNR$, we divide its life span into
  $\Nshell$ epochs. During each epoch, the forward shock propagates into
  the ambient medium and new CRs (and shock-heated ISM plasma) are
  produced. A new spherical shell containing the shocked thermal
  plasma and new CRs is created.  In every subsequent epoch, this
  spherical shell of material evolves while another new shell is
  produced. In this way, an ``onion skin'' structure is formed of shells
  containing CRs of various ages (see Fig.~\ref{Boxel_fig}). The
  evolution of the shells includes the hydrodynamics (i.e., adiabatic
  effects), changes in the assumed frozen-in magnetic field, and losses
  from radiation and \coul\ processes for electrons. 
Spatial diffusion of CRs,
  magnetic field evolution in the shells, and fast synchrotron
  losses for electrons  
  are treated using a finer timescale through further dividing each epoch into 
  a number of time steps $\tstep = 10$\,yr. 
As the local
  magnetic field evolves in the shocked material, the local diffusion
  coefficient is modified accordingly.

As mentioned in footnote~\ref{FN:maxsize}, the maximum CR energy in
our examples is determined by the finite shock size. Particles that
reach this energy escape and, for efficient DSA, carry away a
sizable fraction of the total energy flux.
For each epoch, the CR-hydro simulation determines the escaping flux and
maximum CR energy, $\Emax$, for electrons and protons in the outermost
shell immediately behind the FS where CR acceleration is taking
place. These particles are added to the simulation box in a spherical
shell immediately in front of the FS.  While the precise energy
distribution of the escaped particles is still largely unknown (the
shape is not determined by the CR-hydro model), we assume the escaped
CRs have a Gaussian distribution in momentum-space centered at $\Emax$
and normalized to the total escaped flux ($\Emax$ and the total escaped
flux are determined by the CR-hydro code) \citep{ZP2008}. 
The width of the Gaussian is
determined by fitting the high-energy spectral cut-off around $\Emax$ of
the newly accelerated CRs. The width of this cut-off depends on our
model parameter, $\alfcut$.

As time progresses, the energetic electrons and protons diffuse in both
the SNR shells and in the external material with momentum-dependent
diffusion coefficients described in the next section. As the CRs
diffuse, they interact with the astrophysical environment, such as the
cosmic microwave background (CMB) radiation or a
molecular cloud, and the photon emissivity is recorded as a 3-D map for
later analysis.

\section{Diffusion and Interaction Processes}
\label{section:Diff_Interaction}

\subsection{Diffusion}  % Fig 2
\label{section:diff}

In the simulation, particles spatially diffuse in two distinct regions;
the volume inside the shocked SNR shells and the ambient ISM
outside of the FS. For the shocked material, we assume Bohm diffusion
while for the unshocked ISM we assume much weaker diffusion. For
  this study, we assume Kolmogorov turbulence dominates outside of the
  SNR but other forms could be used instead. 
Inside and outside of the SNR, we assume the
turbulence is strong enough to ensure isotropic diffusion over length
scales $<1$\,pc.

If $\lambda$ is the scattering mean free path, the diffusion
  coefficient can be written as:
\begin{equation} \label{eq:Diff_general}
D =\frac{1}{3} \lambda v   
\ ,
\end{equation}
or, if we assume $\lambda$ is proportional to some power of the
gyroradius,
\begin{equation}
D = D_0 \beta \left ( \rg / \rgz \right )^s
\ ,
\end{equation}
where $v$ is the particle speed, $\beta=v/c$, $\rg = pc/(eB)$ is the
gyroradius in cgs units, 
$\rgz$ is some constant reference length, 
$D_0 = \etamfp \rgz c/3$ is a normalization
constant, and $s$ depends on the magnetic turbulence spectrum.

For Bohm diffusion, $\etamfp=s=1$ and
\begin{equation} \label{eq:Diff_Bohm}
D_{B}=\frac{v}{3} \left ( \frac{pc}{e B(\mathbf{r},t)} \right )
\ .
\end{equation}
Bohm diffusion is assumed throughout the shocked gas and the magnetic
  field in a particular shell, $B(\mathbf{r},t)$, depends on the
  location of the shell and its age.
We assume $B(\mathbf{r},t)$ is `frozen-in' and the details of the field
  evolution are given in \citet{EC2005}.

\begin{figure}        % Fig 2
\begin{center}
\epsscale{1.2} \plotone{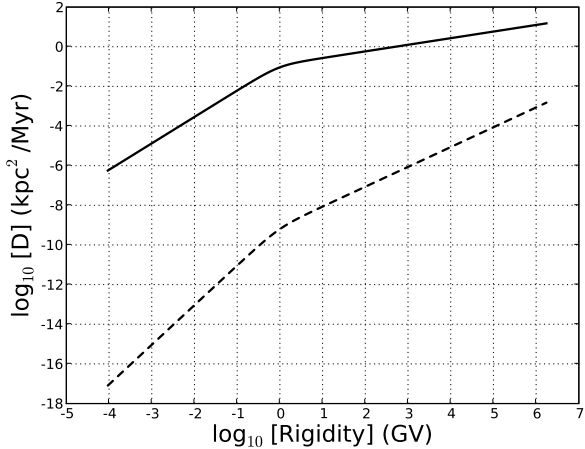}
\caption{ \sloppy Momentum-dependence of spatial diffusion coefficients
for CR protons. Bohm diffusion (dashed curve)  is
implemented for the space inside the SNR shells ($D_{B}$);
The Kolmogorov spectrum (solid curve) is
employed for the space outside the shells for CR diffusion in the
ISM ($\DKol$). A field $B=128$\,\muG\ is
used for the $D_{B}$ plot. The
normalization of $\DKol$ is taken from the calculation for the Galactic ridge by
\citet{PMJSZ2006}.  }
\label{fig:Diff_Kol_Bohm}  
\end{center}
\end{figure}

For the volume outside of the FS, Kolmogorov turbulence
is assumed ($s=1/3$) and the normalization of the diffusion
coefficient $D_\mathrm{Kol}$ is taken from \citet{PMJSZ2006}, 
a value determined to reproduce the observed CR spectra at Earth, i.e.,
\begin{equation} \label{eq:Diff_Kol}
D_\mathrm{Kol} =
0.25 \beta \left ( \frac{R}{10\mathrm{GV}} \right )^{1/3}
\frac{\mathrm{kpc}^{2}}{\mathrm{Myr}}
\ ,
\end{equation}
where $R=pc/(Ze)$ is the magnetic rigidity.
The diffusion coefficients are shown in Fig.~\ref{fig:Diff_Kol_Bohm}
  and, as expected, $D_B \ll D_\mathrm{Kol}$ since the self-generated
  turbulence in the shocked material is far stronger than turbulence in
  the relatively undisturbed ISM.\footnote{We make no attempt to \SCly\
  calculate the turbulence generated by CRs as they escape from the SNR
  and stream through the ISM.}\footnote{We note that 
  Eq.\ref{eq:Diff_Kol} is significantly different from one assumed 
  in a recent paper \citet{Gabici_Aharonian}
  where a strategy to search for ``PeV accelerators" in SNRs is discussed. 
  The difference is due to their assumption that 
  generation of plasma waves can suppress the diffusion coefficient 
  by an order of magnitude relative to that for Galactic cosmic rays.}

Simple diffusion of CR particles is incorporated in the simulation in a
discretized manner.  In each time step and each spatial grid in the 3-D
simulation box, particles are exchanged between the adjacent boxels
according to the particle momentum, location, and density
gradient. 
The particle's location determines which diffusion coefficient is used,
  and the simulation resolution is mainly determined by the boxel size
  and time step, $\tstep$, which are user-tunable.

\subsection{Interaction Processes}
The CR interaction processes considered include
synchrotron radiation, bremsstrahlung, inverse-Compton scattering, and
neutral pion decay. Energy changes from adiabatic effects and radiation,
as well as \coul\ energy losses, are also included.  All of these
processes are treated in a fully space- and time-dependent fashion where
the evolution of relevant parameters, such as the magnetic field and
shell densities, are taken into account in each time step and boxel.

The details of the radiation processes can be found in \citet{Sturner97}
and \citet{BaringEtal99} but we note that, for IC emission, we only
consider CR electrons colliding with a monoenergetic and isotropic
photon field with an average energy density equal to that of
the CMB field.

For hadronic interactions we employ the latest parametric proton-proton
(p-p) model developed by \citet{Kamae06}. In this model, the total
inclusive inelastic p-p cross-section includes the non-diffractive (with
Feymann scaling violation) and diffractive components, plus the
$\Delta(1232)$ and $Res(1600)$ resonance-excitation contributions
important in the 10\,MeV to 1\,GeV range. This model alone can account
for $\sim 20\%$ of the GeV $\gamma$ ray excess between the EGRET
Galactic diffuse spectrum and previous model prediction using proton
data in the solar system \citep{Hunter97}.

\subsubsection{Coulomb Losses}
\begin{figure}       % Fig 3
\begin{center}
\epsscale{1.1} \plotone{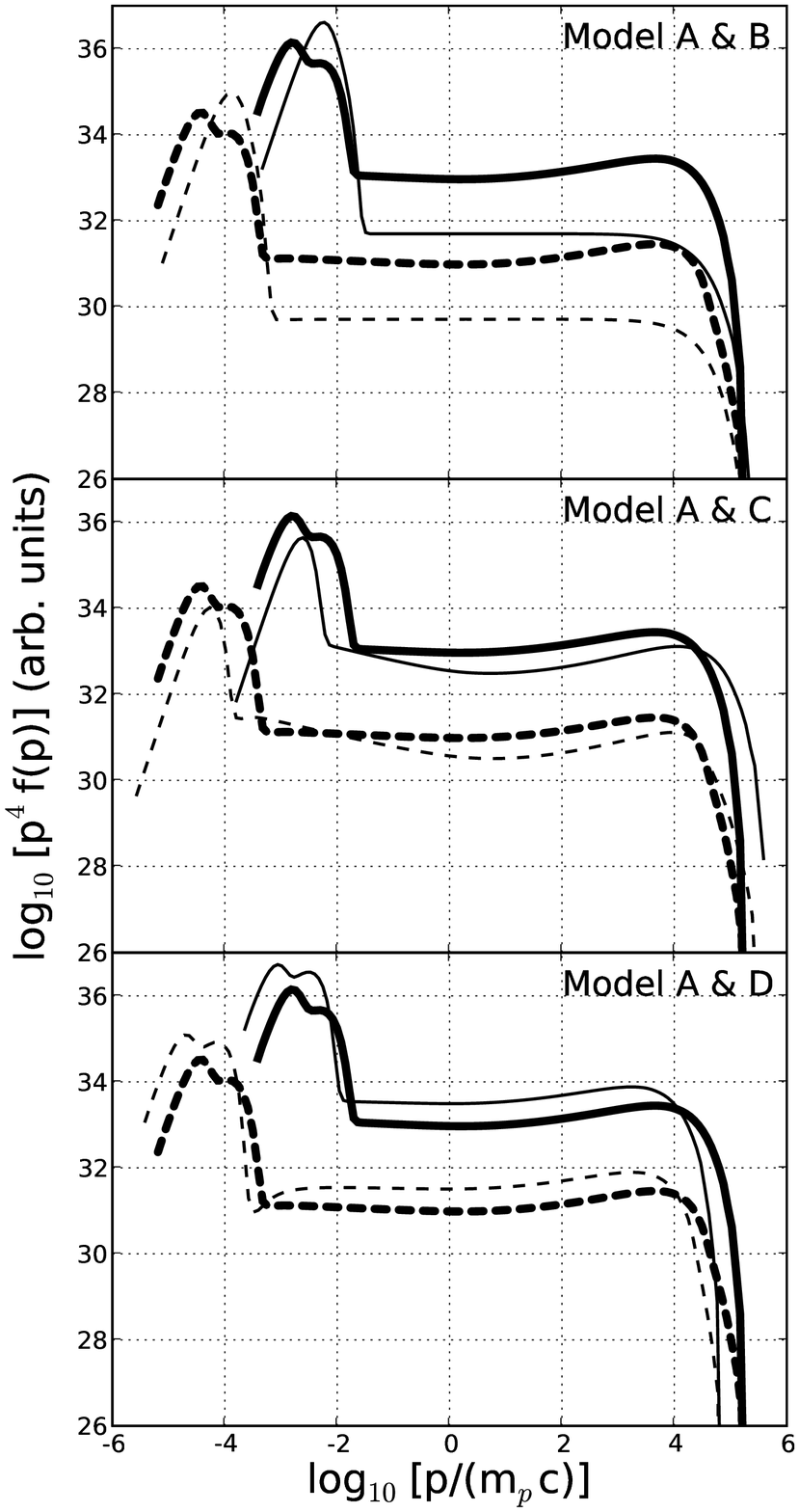}
\caption
{ \sloppy CR spectra integrated over the whole SNR FS-CD region at
    1000\,yr, plotted as $p^4f(p)$, where f(p) is the phase space
    distribution function. In all panels, solid curves are protons and
    dashed curves are electrons. Also in all panels, the heavy curves
    are from Model A and these are compared to thin curves for Model B
    (top panel), Model C (middle panel), and Model D (bottom
    panel). Parameters for the various models are given in
    Table~\ref{table:models}}
\label{fig:total_all_stack}
\end{center}
\end{figure}

Coulomb losses for superthermal electrons are included in our
  model using equation~(10) from \citet{Sturner97}, i.e.,
\begin{equation}
\label{coul}
\dot{E}_\mathrm{coul} = -\left ( \frac{4 \pi e^4}{m_e c}\right ) 
\left [
\frac{\lambda(t) n_\mathrm{SNR} \eta^e_\mathrm{He}}{\beta_e} \right ]
[\psi(t) - \psi'(t)]
\ ,
\end{equation}
where $n_\mathrm{SNR}$ is the proton number density in a shocked shell,
  $\beta_e = v_e/c$ is the electron $\beta$, and $t$ is the time. The
  definitions of the other terms are in \citet{Sturner97} but are not
  important for our discussion here. Equation~(\ref{coul}) shows that
  Coulomb losses increase for large ambient densities and low electron
  speeds. As a shell of shocked material ages, Coulomb losses cause the
  low energy part of the superthermal electron distribution to become
  depleted, as indicated in Fig.~\ref{fig:total_all_stack}. In all
  cases, Coulomb losses are insignifcant for protons.
  
\section{Results}
\label{section:results}
In this initial presentation of our \hbox{3-D} simulation, we show
results for a set of generic Type Ia SNR models where we vary the
acceleration efficiency for DSA, $\EffRel=\EffCRs$, and the
ambient proton number density, $\nH$ (we assume the ISM is made of
hydrogen).  All of the other environmental and model parameters are kept
constant with the values given in Sections~\ref{section:param_envir} and
\ref{section:param_model}.

%Table 1    % tttt
\begin{deluxetable*}{cccccccccc}
\tablecolumns{10}
\tablewidth{6.5in}
\tablecaption{SNR Model Parameters \& Results}
\tablehead{
\colhead{Model}	& \colhead{$\EffRel$}	& \colhead{$\Inj$}	& \colhead{$n_p$}	& \colhead{$\RadFS$\tablenotemark{a}}	& 
\colhead{$\Rtot$\tablenotemark{b}}	& \colhead{$\pmax$\tablenotemark{c}} & \colhead{$\Eesc$\tablenotemark{d}}  & 
\colhead{$F_{\mathrm{keV/TeV}}$\tablenotemark{e}}	& \colhead{$F_{\mathrm{keV/TeV,MC}}$\tablenotemark{f}}	\\
\colhead{} & \colhead{} & \colhead{} & \colhead{(cm$^{-3}$)} & \colhead{(pc)} & \colhead{} & \colhead{($10^{4}m_{p}c$)} & \colhead{} & 
\colhead{}	& \colhead{} 
}
\startdata
A	& 0.75	& 3.70	& 1.0	& 4.67	& 10.66	& 1.37	& 0.27		& 1.24	& 0.97 \\
B	& 0.01	& 4.27	& 1.0	& 5.14	& 4.04	& 1.77	& 0.0015	& 74.0	& 68.5 \\
C	& 0.75	& 3.43	& 0.1	& 6.73	& 10.35	& 3.29	& 0.25		& 5.66	& 4.59 \\
D   & 0.75	& 3.77	& 10.0	& 3.06	& 11.89	& 0.53	& 0.28	 	& 0.19	& 0.16 \\
\enddata
\tablenotetext{a}{Radius of the forward shock at the end of the simulation}
\tablenotetext{b}{Total compression ratio at the end of the simulation}
\tablenotetext{c}{The maximum momentum of CR protons at the end of the simulation}
\tablenotetext{d}{Fraction of $\EnSN$ carried away by the escaped
protons at the end of the simulation. Note that $\EffRel$ includes this fraction.}
\tablenotetext{e}{Energy flux ratio between emission at energies of $3$\,keV and $1$\,TeV at the end of the simulation. The fluxes are integrated over energy bands with widths of $1/10$ of the central energies, and over the entire source volume.}
\tablenotetext{f}{Energy flux ratio as above, but now also including emission from the shell molecular cloud described 
in section~\ref{section:MC_shell}.}  
\label{table:models}
\end{deluxetable*}

The values of $\EffRel$ and $\nH$ for the four
models we compute are given in 
Table \ref{table:models}. We have included the injection efficiency,
$\Inj$, in Table~\ref{table:models} where $\Inj$ determines $\EffCRs$
and in practice we vary $\Inj$ until we obtain the desired acceleration
efficiency.  We also show the fraction of $\EnSN$ 
that is in escaping particles, $\Eesc$.
Model A is used as a reference for the other three and, for all of the
models, the duration of each epoch is 50\,yr so we have 20 shells in the
SNR when it is 1000\,yr old.

\subsection{Electron and Proton Spectra}
In Fig.~\ref{fig:total_all_stack} we show electron and proton
phase-space distributions for the four models listed in
Table~\ref{table:models}.
We plot $p^4f(p)$ to emphasize the spectral
curvature at \rel\ energies and the spectra are integrated over the
entire shocked region between the FS and CD at the end of the
simulation. 

In the top panel, we compare Models A (efficient NL-DSA; bold lines) and
B (inefficient DSA or TP acceleration; thin lines).  The \TP\ model
shows flat electron and proton spectra at \rel\ energies [$f(p)\propto
p^{-4}$] with considerably lower fluxes at \rel\ energies than Model
A. The `thermal' portions of the spectra show that the TP shock produces
higher temperatures than in Model A, a characteristic feature of NL-DSA.
The structure seen in the `thermal' portions of the spectra comes
about because these spectra are summed over the various shells and
the ones produced early on have less efficient DSA and have a
higher temperature than later shells.

\newlistroman
In the middle and lower panels of Fig.~\ref{fig:total_all_stack} we keep
$\EffRel=0.75$ but vary $\nH$; $\nH=0.1$\,\pcc\ in the middle panel and
$\nH=10$\,\pcc\ in the lower panel. The important points for this
comparison are: 
\listromanDE the CR flux at \rel\ energies scales approximately as
$\nH$, as expected,
\listromanDE the maximum proton momentum scales inversely as $\nH$
\citep[see, for example,][]{BaringEtal99},
\listromanDE the electron cutoff energy also scales inversely as $\nH$
but is influenced by radiation losses and the dependence is
weaker than for protons, and
\listromanDE the shocked temperature scales inversely as $\nH$,
although this is not immediately clear from the figures since the
`thermal' portions of the distributions are made up of contributions
from a range of temperatures and densities.

Of course, other aspects of the hydrodynamics depend strongly on
$\nH$. The radius of the SNR at $\tSNR=1000$\,yr is considerably greater
for $\nH=0.1$\,\pcc\ ($\RadFS=6.7$\,pc) than for
$\nH=10$\,\pcc\ ($\RadFS=3.1$\,pc).
It is also expected that the FS will weaken faster with time 
for a denser upstream medium. However the strength, in terms of the
efficiency of NL-DSA, also depends on the magnetic field and 
for the parameters used here, Model D has a larger compression ratio at
$\tSNR=1000$\,yr ($\Rtot=11.9$) than Model A ($\Rtot = 10.7$).  

Coulomb losses also increase as $\nH$ increases and the dip which
appears just above the thermal peak in the Model D electron spectrum
(light dashed curve in lower panel) reflects Coulomb losses experienced
by the superthermal electrons as they collide with the shocked thermal
gas. Coulomb losses can be expected to be more pronounced in NL models
because the larger compression ratio results in a larger post-shock
density. 

\subsection{Spatial Variation and Escaping Flux}

\begin{figure}       % Fig 4
\begin{center}
\epsscale{1.2} \plotone{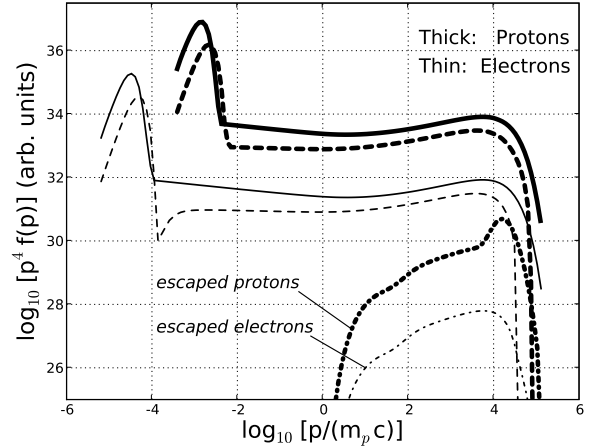}
\caption
{\sloppy 
Spatial variation of CR momentum distribution for Model A at t =
1000yr. The distribution is plotted at three locations: (i) just
behind the FS (solid); (ii) mid-way between the FS and the CD (dashed);
and (iii) at a distance of 9pc away from the SNR center
(dash-dot). Thick lines are for protons and thin lines are for
electrons.  }
\label{fig:spatial_A}
\end{center}
\end{figure}

At any given time, the spatial variation of the CR spectrum can be
calculated. Fig.~\ref{fig:spatial_A} shows CR spectra of Model A at
three different locations: (i) just behind the forward shock (solid
lines); (ii) mid-way between the forward shock and the contact
discontinuity (dashed lines); and (iii) at a distance of $d=9$\,pc from
the center of the SNR which is approximately $2 - 6$\,pc
beyond the FS, depending on the model (dotted lines).
The heavy-weight curves are protons and the light-weight curves are
electrons.

Compared with the freshly accelerated electrons at location (i), many of
the highest energy electrons are lost in the mid-point location (ii)
mainly due to synchrotron losses with a small contribution from
adiabatic losses.  The protons show a smaller change which is due to
adiabatic losses only.

At location (iii), only those CRs that escaped from the shock are
present and their spectra lack a low energy component since low
energy CRs remain trapped in the remnant. The hardness of the spectra at
9\,pc from the center of the SNR reflects the strong momentum dependence
of the escape probability and the spatial diffusion coefficients. The
escape probability from the SNR increases with energy and high-energy
CRs diffuse faster in the ISM.

\subsection{Boardband Photon Spectrum}
Once the particle spectra are determined, the photon emission can be
calculated throughout the simulation box for arbitrary 3-dimensional
distributions of matter and ambient photon fields.

We consider two simple matter distributions (i.e., `molecular clouds')
 outside of the FS.  The first is a spherical shell,
 concentric with the SNR where the inner and outer radii are equal
 to 9 and 10\,pc, respectively. The second is a hemisphere centered at one side
 of the simulation box with radius = 3.2\,pc (see Fig.~\ref{Boxel_fig}).
In both cases, the proton number density in the `molecular cloud' is
$\nMC=10^3$\,\pcc\ and the magnetic field is $\Bism=3$\,\muG, the
same field as in the ISM.
The entire simulation box is 20\,pc on a side and is divided into
$51\times51\times51$ boxels.
The density in the ISM between the molecular clouds and the FS is $\nH$
and the photon field throughout
the simulation box is the uniform CMB field for all models.

\subsubsection{Emission from the SNR Shells}
\label{section:shell}

\begin{figure}       % Fig 5
\begin{center}
\epsscale{1.1}
\plotone{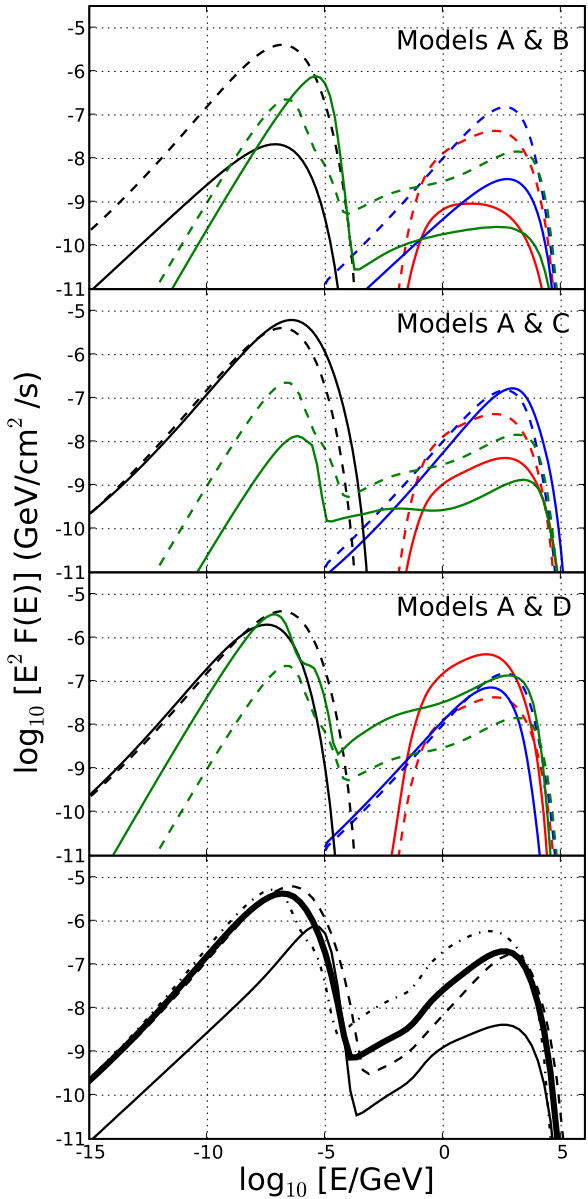}
\caption{%\sloppy
Photon spectra of all four models integrated over the
region from the CD to the FS. Upper three panels:
Models B to D are compared to Model A and
are split into individual components for different emission
mechanisms: $\pi^0$-decay (red), IC (blue),
bremsstrahlung (green) and synchrotron radiation (black). Solid lines
represent spectra for Model B, C and D in each panel while Model A is
shown as dashed lines.  Bottom panel: The contributions from all
mechanisms are summed for each
model: Model A (thick solid), Model B (thin solid), Model C (dash) and
Model D (dash-dot).  
%This array shows how the relative dominance of
%emission mechanisms varies with the model parameters, namely
%acceleration efficiency (Model A vs B) and ambient particle densities
%(Model A vs C and D), in different energy bands.
}
\label{fig:shell_total_comp}
\end{center}
\end{figure}

Fig.~\ref{fig:shell_total_comp} shows the broadband
photon emission for Models A to D integrated over the shocked SNR shells
between the FS and CD. The bottom panel shows the total
spectra while the upper three panels show the
individual components from $\pi^0$-decay (solid), IC (dashed),
synchrotron (dash-dotted), and bremsstrahlung (dotted) compared with
Model A.  Emission from CRs outside of the SNR is not shown.

In Model A, the photon flux in the radio to X-ray energy range is
dominated by synchrotron emission up to $\sim 100$\,keV.  The second
largest contribution is from thermal bremsstrahlung which dominates 
between $\sim 100$\,keV and $\sim 50$\,MeV. 
Between $\sim 50$\,MeV and $\sim 10$\,GeV, \pion\ and IC compete.
Beyond $\sim 10$\,GeV, the emission is dominated by IC.

As seen in the A vs. B comparison panel, thermal bremsstrahlung
plays an important role in the TP Model B and dominates \syn\ emission
in the entire X-ray energy band.  Thermal bremsstrahlung is strong in
the TP model because the shocked temperatures are considerably higher
than those in efficient DSA.  The emission from \syn, IC, and
\pion\ are all weak in the TP case as expected.

In the three NL Models A, C, and D, the acceleration efficiency is set
at $\EffRel=0.75$, but the ambient density
is varied with $\nH =1$, $0.1$ and $10$\,\pcc\ respectively.
In the X-ray band, the thermal bremsstrahlung scales approximately as
$\nH^2$ and dominates \syn\ in Model D, where
$\nH=10$\,\pcc.

Above $\sim 100$\,MeV, the competition is mainly between IC and \pion\
but \brem\ is also important for $\nH=10$\,\pcc. For Model C ($\nH=0.1$\,\pcc), 
both \pion\ and \brems\ are suppressed relative to
IC. For Model D, \pion\ dominates until near the maximum energies where
\brem\ becomes comparable.  

\subsubsection{Emission from a Shell 
`Molecular Cloud'}
\label{section:MC_shell}
We first consider the shell of external material centered with the SNR.
Protons and electrons which have sufficiently high energy and,
therefore, long diffusion lengths can escape from the FS and enter the ISM.
These CRs also interact with the ambient ISM material of density $\nH$
and the CMB radiation.

\begin{figure}         % Fig 6
\begin{center}
\epsscale{1.1}
\plotone{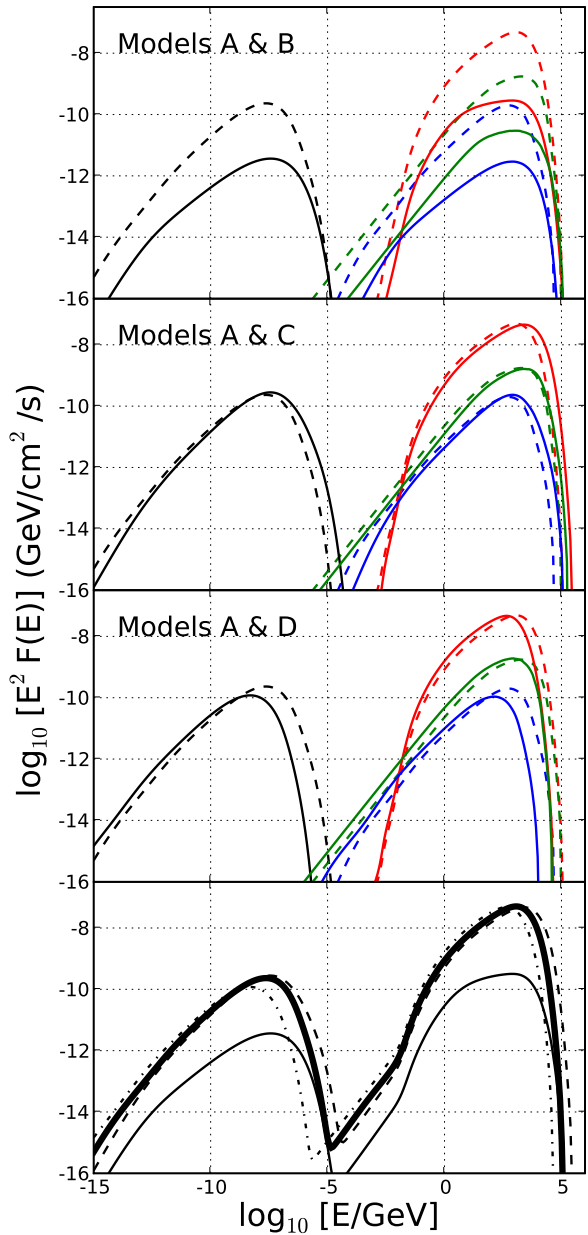} 
\caption
{
\sloppy
Same as Fig.~\ref{fig:shell_total_comp} but the photon spectra
are now integrated over the shell \Mcloud\ volume.
The strong momentum dependent diffusion results
in hard spectra and a change in the relative dominance between the
emission mechanisms, compared to spectra integrated over the SNR.
}
\label{fig:MC_total_comp}
\end{center}
\end{figure}

In Fig.~\ref{fig:MC_total_comp} we show the photon spectra from the
molecular cloud shell for Models A--D in the same representation as
Fig.~\ref{fig:shell_total_comp} but now integrated over the
\Mcloud volume, all calculated at $\tSNR=1000$\,yr.
As expected, these spectra are considerably harder than their
counterparts inside the remnant. 
The escape of CRs from the forward shock during acceleration depends on
 the diffusion coefficient and the strong momentum dependence of the
 Bohm diffusion coefficient, $D_B$, favors the escape of the highest
 energy particles. Once in the ISM, the \rel\ CRs diffuse with a
 diffusion coefficient $D_\mathrm{Kol} \propto p^{1/3}$, hardening the
 spectrum even more, as shown in Fig.~\ref{fig:spatial_A}.
The photon spectra reflect the hard particle spectra.

\begin{figure*}        % Fig 7 
\epsscale{1.0} \plotone{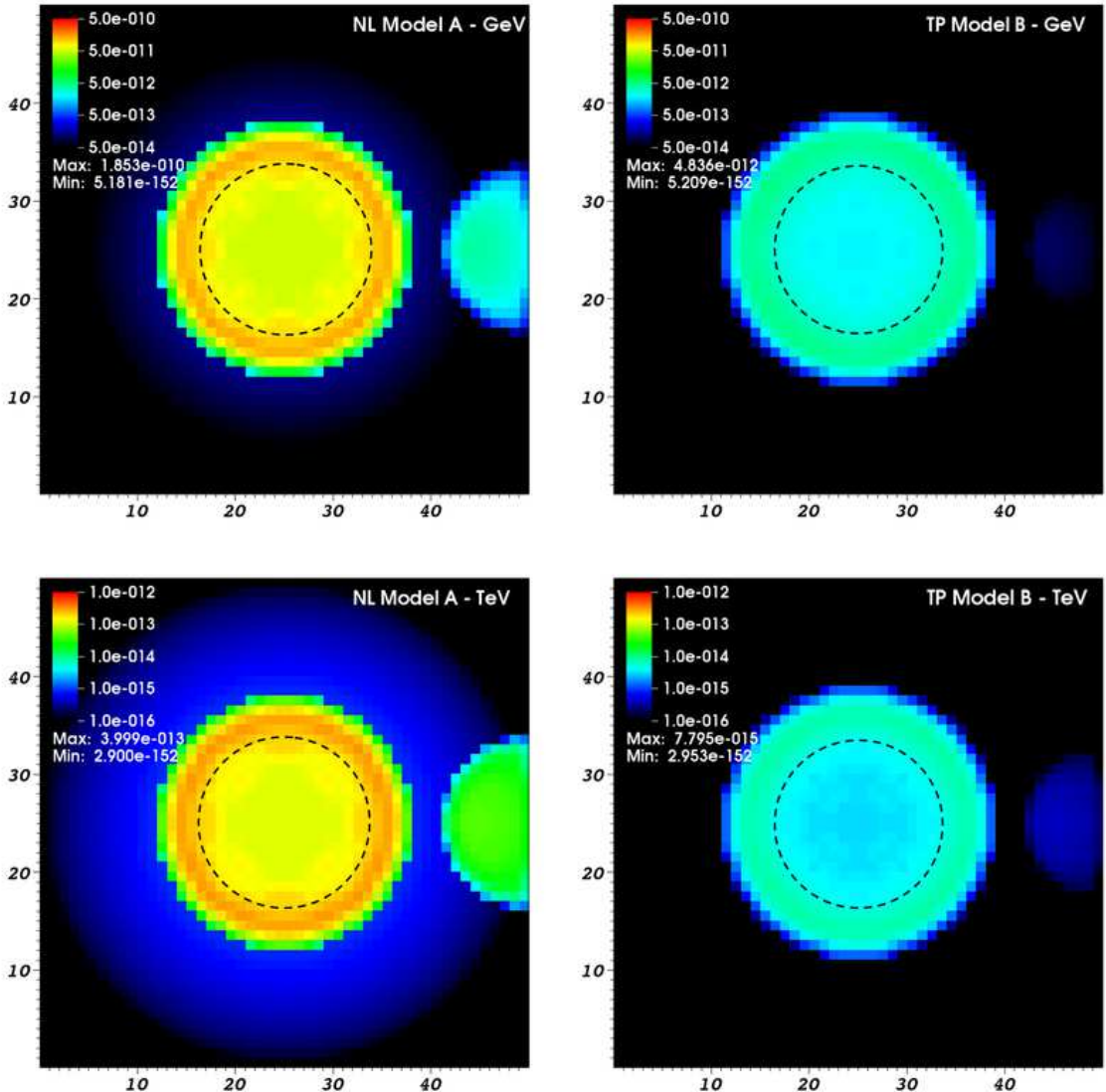}
\caption{ 
Photon flux maps projected along the
line-of-sight for Model A (left panels) and Model B
(right panels), with a hemisphere \Mcloud\
centered at pixel coordinate (50,25,25) with a radius of 3.2\,pc.
The horizontal and virtical scales are in pixels
where the pixel size is 0.38\,pc $\times$ 0.38\,pc.
The upper panels
are integrated over the energy range $1-300$\,GeV, while
the bottom panels are integrated over energies
$E_{\gamma} \ge 1$\,TeV. The color scale is logarithmic in
$\log(N_{\gamma}/\mathrm{cm}^2/\mathrm{s})$.
The dashed circle in each panel indicates the position of the contact discontinuity. 
\label{fig:maps}}
\end{figure*}

\begin{figure}        % Fig 8
\epsscale{1.1} \plotone{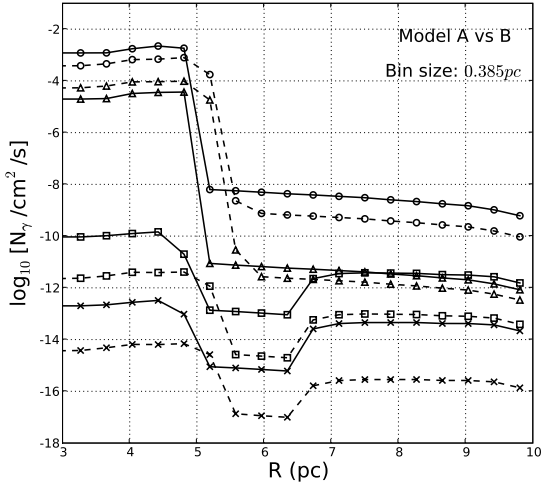}
\caption{
Line-of-sight emission profiles in the radial
direction from the SNR center at $R=0$ to the hemispheric
\McloudPP\ centered at $R=10$\,pc. 
Results for Model A (solid curves) and Model B (dashed curves)
are displayed. Markers represent the centers of the cubic spatial bins which are
$0.385$\,pc on a side. Four wavebands
are considered - (i) Soft X-rays from 1 to 5\,keV (circle); (ii) Hard
X-rays from 5 to 10\,keV (triangle); (iii) 1 to 300\,GeV $\gamma$-rays
(square) and (iv) $\gamma$-rays with energy above 1\,TeV
(cross). The vertical axis shows the photon flux
in log($N_{\gamma}$/cm$^2$/s) for each pixel bin.
}
\label{fig:R_profile_model_A_vs_B}
\end{figure}

\begin{figure}         % Fig 9
\begin{center}
\epsscale{1.1} \plotone{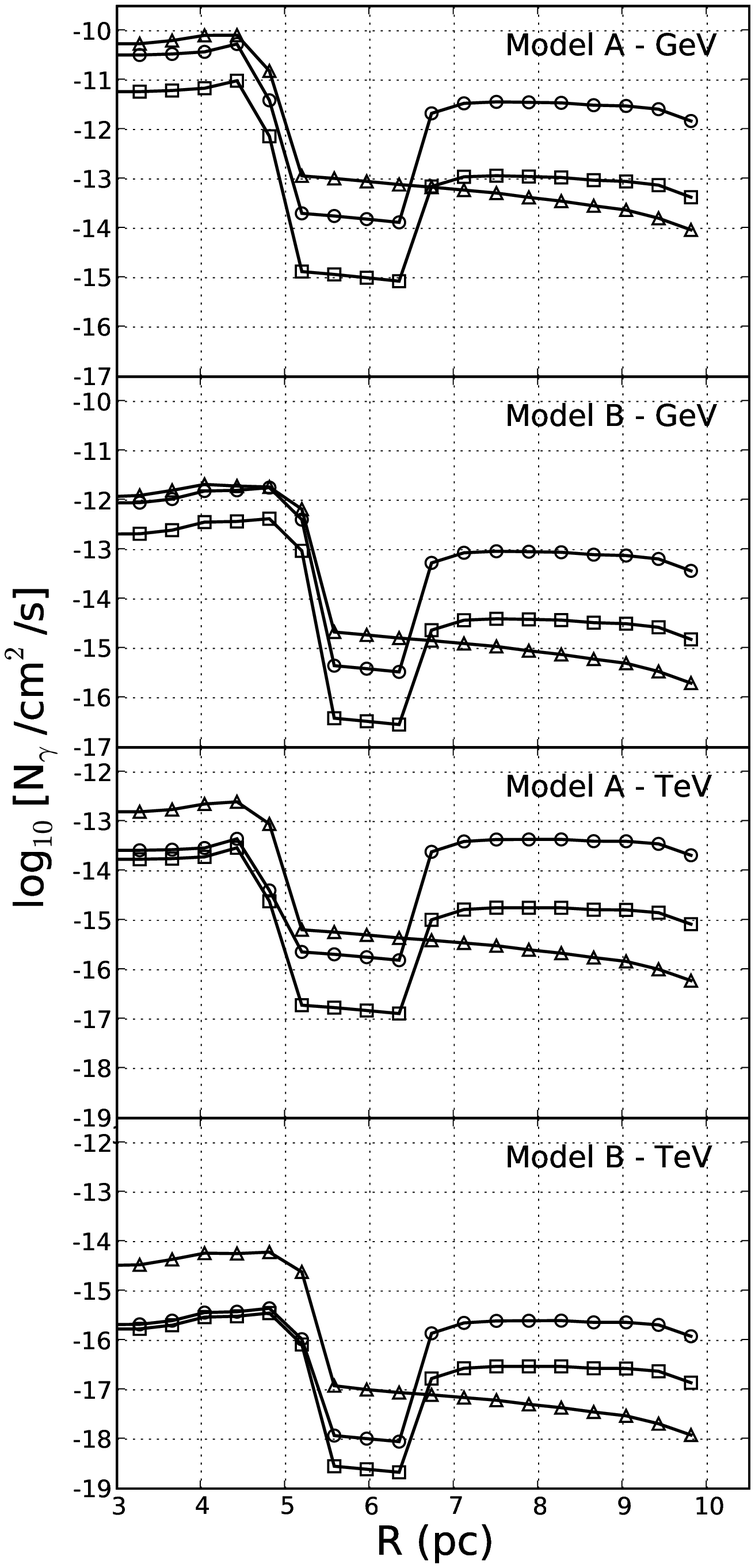} 
\caption{\sloppy Same as Fig.~\ref{fig:R_profile_model_A_vs_B}
    with individual emission processes shown, i.e., \pion\
    (circles), IC (triangles), and bremsstrahlung (squares).
    Synchrotron emission has a negligible flux in the $\gamma$-ray energy band.}
\label{fig:R_profile_comp}
\end{center}
\end{figure}

With a number density of $\nMC=10^3$\,\pcc, and a column density of
$n_\mathrm{Col} \sim 10^{21-22}$\, cm$^{-2}$ in the `molecular cloud,'
$\pi^0$-decay is the main $\gamma$-ray source for all models, followed
by relativistic bremsstrahlung and then IC emission, as shown in
Fig.~\ref{fig:MC_total_comp}.
For Models A, B, C and D, the separation between the
FS and the inner edge of the MC is found to be around $4.3$, $3.8$,
$2.3$ and $5.9$\,pc respectively at $t=1000$\,yr.

For the environmental parameters studied here, the emission at all
wavebands from the \Mcloud\ is weaker than that from the SNR
shell, but the difference depends on the photon energy.  With the
assumption $B_\mathrm{MC} = 3$\,\muG, the X-ray \syn\ flux from the
\Mcloud\ stays at the ISM level and will be difficult to
detect. The GeV $\gamma$-ray flux is more model-dependent and the
flux stays around a factor of 10--100
smaller than the flux from the SNR.
For the TeV flux, which is detected by atmospheric Cherenkov
telescopes, the \Mcloud\ emission is about $1-10\%$ of that from the
SNR.

\subsection{Boardband Images and Projected Emission Profiles}
\label{section:profile}
Multi-wavelength projection maps are useful for studying the
energy-dependent morphology of SNRs. 
We use our hemispherical \McloudPP\ example (Fig.~\ref{Boxel_fig}) to
calculate 2-D projection maps in various energy bands at
$\tSNR=1000$\,yr for Models A and B.  After the photon emissivity is
calculated in each boxel in the 3-D simulation box, we perform a
line-of-sight projection through the box.
\newlistroman
We choose four energy bands:
\listromanDE soft X-rays with $\Egam = 1-5$\,keV;
\listromanDE hard X-rays with $\Egam = 5-10$\,keV;
\listromanDE $\Egam = 1-300$\,GeV; and
\listromanDE $\Egam >1$\,TeV.  
The parameters we use result in a column density of $\sim
10^{21}$\,cm$^{-2}$ for the cloud, which is small enough to ignore in
the present context.

Fig.~\ref{fig:maps} shows the
$\gamma$-ray projected flux maps in
$\log(N_{\gamma}/\mathrm{cm}^2/\mathrm{s})$ at a source distance of
$\dSNR=1$\,kpc for Models A (NL) and B (TP) in
the GeV and TeV bands [i.e., bands (iii) and (iv)].
The color scales are different for the GeV and TeV images,
but the spatial resolution is the same.
The difference between the example with efficient DSA (left panels)
  and the \TP\ case (right panels) is mainly one of intensity if only
  GeV-TeV emission is concerned. In both cases, the brightest regions of
  the SNR are considerably brighter than the cloud and for the \TP\ case
  (Model B), the cloud is almost invisible on these scales. For the SNR
  in both the NL and TP cases, the region between the CD and FS clearly
  shows up in the maps even with the projection through the
  remnant.\footnote{The dashed circle in each panel shows the
  position of the CD at 1000\,yr.} 
There is also a clear limb darkening effect from projection
  seen at the edge of both remnants and at the edge of the cloud in the
  NL case. At the \Mcloud, however, there is no noticable drop in
  intensity towards the center of the cloud as occurs for the SNR.

These details, of course, depend on the particular parameters we
  have chosen but some general statements can be made. 
  Unless there is a source of soft photons associated with the
  external material, the brightness of the external material (MC)
  compared to the SNR, the $\Iratio$ ratio, will be independent of the
  density ratio $\nMC/\nH$ if IC dominates the GeV-TeV emission. If
  \pion\ or \brem\ dominate, the $\Iratio$ ratio will scale
  approximately as the first power of the density ratio, $\nMC/\nH$.
  In all cases, $\Iratio$ will decrease with the distance the external material is
  from the FS. Another important result, which is implicit in
  Fig.~\ref{fig:maps} and important for comparing \pion\ and IC
  emission, is that emission from the SNR and the external material must
  be considered together. To first order, an increase in acceleration
  efficiency or ambient matter density not solely associated with the
  cloud, $\nH$, will leave $\Iratio$ unchanged.
%The brightness of
%  the external material (MC) compared to the SNR, the $\Iratio$ ratio,
%  will scale approximately as the first power of the density ratio,
%  $\nMC/\nH$, if IC dominates the GeV-TeV emission, and approximately as
%  the second power of $\nMC/\nH$ if \pion\ dominates. 
%

In Fig.~\ref{fig:R_profile_model_A_vs_B} we show emission calculated
along a horizontal line from the center of the remnant at $R=0$
across the \Mcloud\ for all four energy bands. These fluxes are
determined, as are the 2-D maps, by summing the emission from each boxel
along a line-of-sight.
The plateaus on the left hand side of the plot within $R \lesssim 5$\,pc
 show emission from the SNR.  The subtle increase of the projected flux
 with $R$ in this region is the result of projection through the
 shell of material between the CD and the FS. 
Beyond $R\sim 5$\,pc, the fluxes drop abruptly to the ISM level. Here,
escaping CRs stream through the ISM with a large diffusion
coefficient $\DKol$.
At $R \sim 6.8$\,pc, the CRs impact the hemisphere \McloudPP\ with
$\nMC=10^3$\,\pcc\ and the fluxes for energy bins (iii) and (iv)
increase by almost 2 orders of magnitude from the ISM level. 
These photons are mainly from \pion. There is no increase at the
  edge of the cloud for energy bins (i) and (ii) since this emission is
  totally from \syn\ and we have assumed the field in the \Mcloud\
  equals the ISM field, $\Bcloud=\Bism$.

Fig.~\ref{fig:R_profile_comp} shows the emission profiles in bands
(iii) and (iv) for Models A and B separated into individual
emission mechanisms. While the total fluxes at these energies
depend strongly on acceleration efficiency, the $\Iratio$ ratio is much
less sensitive to $\EffRel$,
as mentioned above. There is no increase in IC emission at the edge of
the cloud near $R\sim 6.9$\,pc since we only consider electron
scattering off the CMB. Unless there is an additional source of photons
associated with the external material, IC will be strongly suppressed
relative to \pion\ in external material.

\section{Summary and Conclusions}
\label{section:summary}
We have introduced a 3-D simulation of an evolving SNR where the \NL\
acceleration of CRs is coupled to the SNR evolution.  The model follows
the diffusion and interaction of CRs within the spherically
symmetric remnant, as well as high-energy CRs that escape from the
forward shock and diffuse into the surrounding medium.
For any set of model and environmental parameters, and for arbitrary
distributions of matter surrounding the remnant, we can calculate
broadband photon spectra and obtain line-of-sight projections and
morphologies that will allow for efficient comparisons with observations
in various energy bands.

We have illustrated the capabilities of this simulation with several
models that differ from each other in the CR acceleration efficiency,
$\EffRel$, the ambient ISM proton density, $\nH$, and the matter
distribution of a \McloudPP\ external to the SNR.
Of course, all of the results discussed here assume particular
  values for parameters, such as a shocked
  electron to proton temperature ratio $T_e/T_p = 1$ and an
  electron/proton ratio at \rel\ energies $\epRatio=0.01$.  These
  parameters are critical for understanding DSA and applying the
  mechanism to astrophysical sources yet they are poorly constrained by
  both observations and theory.  For instance, the value of $T_e/T_p$
  determines the importance of \brems\ compared to \syn\ in the X-ray
  range and also strongly influences the thermal X-ray line spectra
  \citep[][]{EPSBG2007}.
The $\epRatio$ ratio is the most important factor after the ambient
density determining the relative intensity of IC and \pion\ emission at
GeV-TeV energies. The confirmation of CR {\it ion} production in SNRs
depends on this parameter.

Other important parameters of DSA that remain uncertain are the
injection and acceleration efficiencies, the amount of magnetic
compression and amplification that occurs, and the diffusion coefficient
of escaping particles as they leave the shock, which must differ
substantially from Bohm diffusion \citep[e.g.,][]{BAC2007,EV2008}.
Due to the still limited dynamic range of particle-in-cell simulations,
and the lack of strong, \NL\ shocks producing \rel\ particles in the
heliosphere, we believe young SNRs are the best `laboratory' for
studying NL-DSA. Broadband observations matched against \SC\ \NL\ models
currently provide the best constraints on these important
parameters.

There are three important aspects of our 3-D
simulation that are new and extend the large body of existing work on
DSA in SNRs.
One is that the simulation consistently models high-energy CRs that
escape from the forward shock of the SNR with the evolution of the SNR
itself. In NL-DSA, the fraction of total explosion energy that ends up
in escaping particles can be large (see Table~\ref{table:models}) and we
believe this is the first work to include these particles in a coherent
emission model.
Second, the 3-D simulation box
allows for the modeling of CR interactions in arbitrary mass
distributions outside of the SNR. This feature is essential for
producing 2-D projection maps that can be compared with current and
future observations. 
These maps, tuned to match the instrument response of telescopes, will
 serve to help determine the importance of pre-SN shells and/or nearby
 molecular clouds in producing \gamray\ emission.
Third, the simulation platform is extremely flexible making it
straightforward to add important effects not present in this preliminary
model.  These generalizations include shock acceleration and heating at
the reverse shock as well as the forward shock, pre-SN winds for Type II
SNe, various forms for particle diffusion in the ISM, 
production and interaction of heavy CR ions, and a parameterized representation of
magnetic field amplification.

Another physical effect that may importantly influence the photon
spectrum is anisotropy from angular-dependent interactions. These
include an angular-dependent neutral pion production cross-section
\citep{Karlsson07} and anisotropic IC scattering with photon fields
other than the CMB \citep{Moskalenko00}. Preliminary results show that
anisotropies can change the spectral shape and flux of the observed
photons drastically. 
When anisotropic interactions are implemented, the projection maps we
calculate will show how the observed flux depends on the orientation of
the FS and \Mcloud\ with respect to the line-of-sight. We leave this
issue to future studies.

Finally, in addition to modeling the photon emission from SNRs, our
model can also determine the total contribution of CR ions and electrons
injected into the Galaxy from an individual SNe over its lifetime.  This
can serve as input to Galaxy-scale propagation models \citep[for
example, GALPROP,][]{SMP2007} and also add to our knowledge on the
Galactic \gamray\ diffuse emission.

\acknowledgments
The authors wish to thank Roger Blandford,
Steven Kahn, Igor Moskalenko, Niklas Karlsson, Stefan Funk, Takaaki Tanaka, Johan-Cohen 
Tanugi and Masaru Ueno for helpful discussions. 
They are grateful to the anonymous referee for bringing new
publications to their attention.
D.C.E. is grateful for the hospitality 
of KIPAC where part of this work
was done, as well as for support from a NASA ATP grant (06-ATP06-21) and
a NASA LTSA grant (NNH04Zss001N-LTSA). This work was supported in part
by the U.S. Department of Energy under Grant DE-AC02-76SF00515.

\end{document}